\documentstyle{aipproc}
 \newcommand{\be}{\begin{equation}}
 \newcommand{\ee}{\end{equation}}
 \newcommand{\ba}{\begin{eqnarray}}
 \newcommand{\ea}{\end{eqnarray}}
 \def\entry#1#2{\vbox{\hbox to 100truept{\hrulefill}\break%
                \hbox{\vrule\vbox to 28truept{%
                \vfill%
                \hbox to 100truept{\hfill\quad\small #1\quad\hfill}\break%
                \vfill%
                \hbox to 100truept{\hfill\quad\small #2\quad\hfill}%
                \break\vfill%
                \hbox to 100truept{\hrulefill}}\vrule}}}%

 \def\arrwv#1#2{\vbox to 42truept{\vfill%
               \hbox to 92truept{\put(48,0){\line(0,-1){10}}\hfill}\break%
                \vfill%
                \hbox to 92truept{\hfill\small #1\hfill}\break%
                \vfill%
                \hbox to 92truept{\hfill\small #2\hfill}\break%
                \hbox to 92truept{%
                \put(48,0){\vector(0,-1){10}}\hfill}\break
                \vfill}}

 \def\arrwh#1#2{\vbox to 28truept{\vfill
                \hbox to 92truept{\small\hfill #1\hfill}\break
                \hbox to 92truept{\rightarrowfill}\break
                \hbox to 92truept{\small\hfill #2\hfill}\break\vfill}}

 \def\d{\partial}
 \def\real{{\vrule height 1.0ex
             width 0.05em depth 0ex \kern -0.06em {\rm R}}}
 \def\Journal#1#2#3#4{{#1} {\bf #2}, #3 (#4)}

 \def\ANP{\it Ann.\ Physics (N.Y.) }

 \def\IJMPA{\it Int.\ J.\ Mod.\ Phys.\ A }
 \def\IJMPD{\it Int.\ J.\ Mod.\ Phys.\ D }
 \def\MPLA{\it Mod.\ Phys.\ Lett.\ A }
 
 \def\NPBP{\it Nucl.\ Phys.\ (Proc.\ Suppl.) }
 \def\PLB{\it Phys.\ Lett.\  B }
 
 \def\PRD{\it Phys.\ Rev.\ D }
\begin{document}
\title{Two-Dimensional Dilaton Gravity}
\author{Marco Cavagli\`a}
\address{Max-Planck-Institut f\"ur Gravitationsphysik,
Albert-Einstein-Institut,\\
Schlaatzweg 1, D-14473 Potsdam, Germany
\thanks{E-mail:
cavaglia@aei-potsdam.mpg.de, \hbox to 7truecm{\hfill}
web page: http://www.aei-potsdam.mpg.de/\~{}cavaglia}}
\maketitle
\begin{abstract}
I briefly summarize recent results on classical and quantum
dilaton gravity in 1+1 dimensions.
\end{abstract}

\section{Introduction}\label{introduction}

In the last few years a great deal of activity has been devoted to the
investigation of lower-dimensional gravity \cite{Web}. The interest on
dimensionally reduced theories of gravity relies essentially on their
relation to string theory, higher-dimensional gravity, black hole physics,
and gravitational collapse. In this talk I will focus attention on the
simplest, non-trivial, lower-dimensional theory of gravity: 1+1 (pure)
dilaton gravity \cite{matter}. 
 
Dilaton gravity is described by the action
\be
S[\phi,g_{\mu\nu}]=\int d^2x\sqrt{-g}[\phi R(g)+V(\phi)]\,,\label{action-gen} 
\ee
where $\phi$ is the dilaton field, $V(\phi)$ is the dilatonic potential,
and $R$ is the two-dimensional Ricci scalar. In Eq.\ (\ref{action-gen}) we
have used a Weyl-rescaling of the metric to eliminate the kinetic
term of the dilaton field.  Equation (\ref{action-gen}) describes a
family of models whose elements are identified by the dilatonic potential.
For instance, $V(\phi)=constant$ identifies the matterless sector of the
Callan-Giddings-Harvey-Strominger model (CGHS) \cite{CGHS}, $V(\phi)=\phi$
identifies the Jackiw-Teitelboim model, and $V(\phi)=2/\sqrt{\phi}$
describes the two-dimensional sector of the four-dimensional
spherically-symmetric Einstein gravity after having integrated on the
two-sphere with area $4\pi\phi$. 

Dilaton gravity is an interesting example of {\em Completely Integrable
Model}, i.e.\ a model that can be expressed in terms of free fields by a
canonical transformation. Completely integrable models play an important
role from the point of view of the quantum theory because they can be
quantized exactly (in the free-field representation). This property allows
the discussion of quantization subtleties and non-perturbative quantum
effects. (For the CGHS model see for instance Refs.\ \cite{Jackiw}.) 
Since dilaton gravity can be used to describe black holes and/or
gravitational collapse (in the case of coupling with matter), the
quantization program is worth exploring. 

A direct consequence of the complete integrability of dilaton gravity
is that both the metric and the dilaton can be expressed in terms of a
D'Alembert field and of a local integral of motion independent of the
coordinates \cite{Filippov}. So, using the gauge in which the free field
is one of the coordinates, one finds that all solutions depend on a single
coordinate. This property constitutes a generalization of the classical
Birkhoff Theorem. (For spherically-symmetric Einstein gravity, i.e.\
$V(\phi)=2/\sqrt{\phi}$, the ``local integral of motion independent of the
coordinates'' is just the Schwarzschild mass and the dependence of both
the metric and the dilaton from a single D'Alembert field means that the
four-dimensional line element can be written in a form depending on the
radial coordinate only.) 

So dilaton gravity can be quantized using two alternative, a priori
non-equivalent, approaches. In the first one the theory is quantized by
first reducing it to a 0+1 dynamical system, i.e.\ using first the
classical Birkhoff theorem and then the quantization algorithm.
Conversely, in the second approach the theory is quantized in the full 1+1
sector and the 0+1 dimensional nature of the system must be recovered a
posteriori ({\em Quantum Birkhoff Theorem}) \cite{Birkhoff}: 
\be
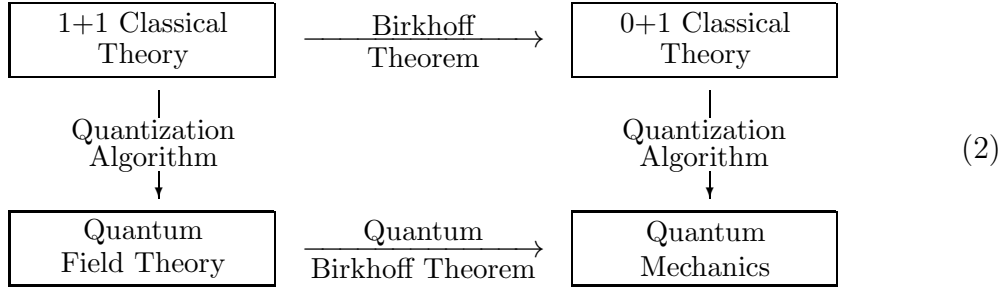
\begin{array}{rcc}
\entry{1+1 Classical}{Theory}
&\arrwh{Birkhoff}{Theorem}&\entry{0+1 Classical}{Theory}\\
\arrwv{Quantization}{Algorithm}&&\arrwv{Quantization}{Algorithm}\\
\entry{Quantum}{Field Theory}&\arrwh{Quantum}{Birkhoff Theorem}&
\entry{Quantum}{Mechanics}
\end{array}
\ee
Furthermore, because of the gauge nature of the theory, the quantization
of the system can be implemented according to two different procedures:
the {\em Dirac method} -- quantization of the constraints followed by
gauge fixing -- and the {\em reduced canonical method} -- classical gauge
fixing followed by quantization in the reduced space. Usually, the two
methods do not lead to identical results.
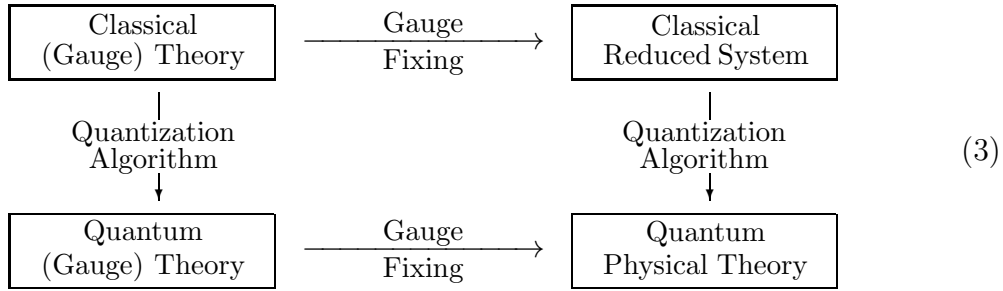
\be
\begin{array}{rcc}
\entry{Classical}{(Gauge) Theory}
&\arrwh{Gauge}{Fixing}&\entry{Classical}{Reduced System}\\
\arrwv{Quantization}{Algorithm}&&\arrwv{Quantization}{Algorithm}\\
\entry{Quantum}{(Gauge) Theory}&\arrwh{Gauge}{Fixing}&
\entry{Quantum}{Physical Theory}
\end{array}
\ee
Here I will show that both diagrams close and the different approaches are
equivalent. At the end of the talk I will briefly discuss why these
conclusions fail in the case of dilaton gravity coupled to scalar matter. 

\section{0+1 Quantization}\label{0+1}

In the 0+1 approach the proof of the equivalence of Dirac and reduced
methods is straightforward because we are able to pass, via a canonical
transformation, to a maximal set of gauge-invariant canonical variables. 

Using obvious notations the 0+1 action reads
\be
S_{0+1}=\int dt[\dot q_i p_i-\mu H]\,,~~~~~i=1,2\,,
\ee
where $\mu$ is a Lagrange multiplier enforcing the constraint $H=0$. Thus
in the 0+1 sector of the theory we can express the field equations as a
canonical system in a finite, $2\times 2$ dimensional, phase space. 

Clearly, due to the complete integrability of the model, the equations of
motion are analytically integrable and their solution coincides with the
finite gauge transformation generated by the (single) constraint $H=0$. So
we can find a couple of gauge-invariant independent canonical quantities
($M,P_M$) and construct the maximal gauge-invariant canonical chart
$(M,P_M,H,T)$. Now $T$ can be used to fix the gauge because its
transformation properties for the gauge transformation imply that time
defined by this variable covers once and only once the symplectic
manifold, i.e.\ time defined by $T$ is a global time. The quantization
becomes trivial and both Dirac and reduced approaches lead to the same
Hilbert space. The Hilbert space is spanned by the eigenvectors of the
(gauge invariant) operator $M$ corresponding to the ``mass'' of the system
introduced in the previous section.

The quantization program illustrated above has been implemented in detail
in Refs.\ \cite{bh} for the case of spherically-symmetric Einstein gravity
but can be easily generalized to an arbitrary $V(\phi)$. In the case of
Refs.\ \cite{bh} one can go further and prove that the Hermitian operator
$M$ in the gauge fixed, positive norm, Hilbert space is not self-adjoint,
while its square is a self-adjoint operator with positive eigenvalues.
This result is due to the fact that the conjugate variable to the ``mass''
$M$, $P_M$, has positive support, analogously to what happens for the
radial momentum in ordinary quantum mechanics. It would be interesting to
explore whether this conclusion holds for other choices of the dilatonic
potential. In any case, what is important for the present discussion is
that the mass $M$ -- or its square -- is the only gauge-invariant
observable of the system (apart from the conjugate variable, of course). 

\section{1+1 Reduced Quantization}\label{reduced}

The reduced quantization of the full 1+1 theory can be implemented using
``geometrodynamical-like'' canonical variables similar to the canonical
variables introduced by Kuch\v{a}r for the canonical description of the
Schwarzschild black hole \cite{Kuchar}. The new variables are directly
related to the spacetime geometry and the relation to the 0+1 formalism is
straightforward.

Let us introduce the ADM parametrization of the metric
\be
g_{\mu\nu}=\rho\left(\matrix{\alpha^2-\beta^2&\beta\cr
\beta&-1\cr}\right)\,,\label{metric}
\ee
where $\alpha(x_0,x_1)$ and $\beta(x_0, x_1)$ play the role of the lapse
function and of the shift vector respectively, and $\rho(x_0,x_1)$
represents the dynamical gravitational degree of freedom. Using Eq.\
(\ref{metric}), the two-dimensional action in the Hamiltonian form reads
\be
S[\phi,g_{\mu\nu}]=\int d^2x\sqrt{-g}[\dot\phi\pi_\phi+\dot\rho\pi_\rho-
\alpha {\cal H}_0-\beta {\cal H}_1]\,.\label{action-hamil}
\ee
We can pass to a new canonical chart
$(M,\pi_M,\bar\phi,\pi_{\bar\phi})$
using the canonical map
\ba
&M=N(\phi)-\displaystyle{\rho^2\pi_\rho^2-\phi'^2\over\rho}\,,~~~
&\pi_M=\displaystyle{\rho^2\pi_\rho\over
\rho^2\pi_\rho^2-\phi'^2}\,,\\
&\bar\phi=\phi\,,~~~~~~~~~~~~~~~~~~~~~~~~~~~
&\pi_{\bar\phi}=\pi_\phi-\displaystyle{\rho^2\pi_\rho\over
\rho^2\pi_\rho^2-\phi'^2}\left[V(\phi)+2\pi_\rho\left({\phi'\over
\rho\pi_\rho}\right)'\right]\,.\label{tr-phi}
\ea
The canonical quantity $M$ corresponds to the local integral of motion
mentioned in Sec.\ \ref{introduction} and can be identified with the mass
of the system. In the 0+1 sector $M$ reduces to the quantity defined in
Sec.\ \ref{0+1}. In the new canonical chart the ADM super-Hamiltonian and
super-momentum constraints read
\be
{\cal
H}_0=[N(\phi)-M]\pi_{\bar\phi}\pi_M+[N(\bar\phi)-M]^{-1}\bar\phi'M'\,,~~~~
{\cal H}_1=-\bar\phi'\pi_{\bar\phi}-M'\pi_M\,,\label{constraints}
\ee
where $'$ means differentiation w.r.t.\ the spatial coordinate $x_1$.

The canonical action (\ref{action-hamil}) must be complemented by a
boundary term at the spatial infinities. This can be done along the lines
of Refs.\ \cite{Kuchar}. The resulting boundary term is of the form
\be
S_\d=-\int dx_0(M_+\alpha_++M_-\alpha_-)\,,\label{boundary}
\ee
where $M_\pm\equiv M(x_0,x_1=\pm\infty)$ and $\alpha_\pm(x_0)$ parametrize
the action at infinities. 

Now we can solve the constraints and quantize the theory. It is easy to
prove that the general solution of Eqs.\ (\ref{constraints}) is given by
\be
\pi_{\bar\phi}=0\,,~~~~~~M'=0\,.\label{sol-constr}
\ee
(Note that $M$ weakly commutes with the constraints, as expected for a
local integral of motion.) Thus $M\equiv m(x_0)$ and the
effective Hamiltonian is simply given by the boundary term
(\ref{boundary}). The reduced action reads
\be
S[m]=\int d\tau\left[\frac{dm}{d\tau} p-m\right]\,,\label{action-red}
\ee
where $p=\int_{\real}dx_1\pi_M$ and $\tau(x_0)=\int^{x_0}
dx_0'(\alpha_++\alpha_-)$. Now quantization can be carried on as
usual. The Schr\"odinger equation is
\be
i{\d\over\d\tau}\psi(m;\tau)=H_{\rm eff}~\psi(m;\tau)\,,~~~~~~~~~H_{\rm
eff}\equiv m\,.
\ee
The stationary states are the eigenfunctions of $m$ and the Hilbert space
coincides with the Hilbert space obtained in the 0+1 approach.

\section{1+1 Dirac Quantization}\label{dirac}

The equivalence between 0+1 and 1+1 Dirac methods can be easily
proved using the canonical transformation illustrated in the previous
section. However, the same result can be obtained through a
completely different quantization scheme. Let me sketch the main points.

Since dilaton gravity is completely integrable \cite{Filippov}, it seems
reasonable to assume the existence of a canonical transformation mapping
the original system to a system described by a pair of (constrained) free
fields $A_\alpha$ ($\alpha=0,1$) in a flat two-dimensional Minkowski
background.  (In the CGHS case, i.e.\ constant dilatonic potential, this
canonical transformation is explicitly known since long time
\cite{Jackiw}. A generalization to linear and exponential dilatonic
potentials has been recently derived by Cruz and Navarro-Salas, see Ref.\
\cite{Navarro}.)  Using the free fields $A_\alpha$ the super-Hamiltonian
and super-momentum constraints read
\ba
&{\cal
H}_0=\displaystyle\frac{1}{2}\pi^\alpha\pi_\alpha+\frac{1}{2}A'^\alpha
A'_\alpha=0\,,\label{Ham-CGHS}\\
&{\cal H}_1=-\pi^\alpha A'_\alpha=0\,,\label{mom-CGHS}
\ea
where $\pi^\alpha$ are the conjugate momenta of $A_\alpha$. Now the theory
can be quantized in the free field representation. This has been done in
detail in Ref.\ \cite{Birkhoff} for the CGHS model. (See also Refs.\ 
\cite{Jackiw}.) Since in this case the canonical transformation is known
explicitly, the equivalence with the previous approaches can be proved.

In the CGHS case the functional $M$ defined in Sec.\ \ref{introduction}
and Sec.\ \ref{reduced} is \cite{Birkhoff}
\be
M=M_0+M_1(A_\alpha,\pi_\alpha)\,,
\ee
where $M_0$ is a constant (zero mode). On the field equations we have
$M=M_0$. Due to positivity conditions that are present in the model the
constraints (\ref{Ham-CGHS}-\ref{mom-CGHS}) can be linearized and the
quantization is carried out by use of the standard Gupta-Bleuler method
\cite{Birkhoff}.

The quantum reduction of the theory to a 0+1 dynamical system can be made
clear by investigating the matrix elements of the operator $M$. 
Adopting a normal ordering the matrix elements of $M$ between physical
states are
\be
<\Psi_2|M|\Psi_1>=<\Psi_2|M_0|\Psi_1>\,.
\ee
Since $M_0$ is a zero mode, it commutes with all the creation and
annihilation operators of $A_\alpha$. So the vacuum must be
labeled by the eigenvalue of $M_0$, i.e.\
\be
M_0|0;m>=m|0;m>\,.
\ee
The existence of infinite vacua, differing by the eigenvalue of the mass,
implies that the theory reduces to quantum mechanics. Again, the only
gauge invariant operator is the mass (and its conjugate momentum) and the
resulting Hilbert space is spanned by the eigenvectors of $M$. 

\section{Coupling to a Massless Scalar Field}

Let me conclude this talk spending few words on dilaton gravity coupled to
a massless scalar field. We have seen that the topological nature of
dilaton gravity is a direct consequence of the existence of a functional
of the canonical variables which is conserved under time and space
translations (the mass $M$): the original fields can be expressed in terms
of a free field and a local integral of motion instead of two free fields,
as one might expect from the counting of the degrees of freedom.

When a scalar field is coupled to the system non-static solutions appear,
the Birkhoff theorem is no longer valid, and the topological nature of
dilaton gravity is destroyed. This has an important consequence from the
canonical point of view. Indeed, we can immediately conclude that no local
integrals of motion like $M$ do exist. A provocative interpretation of
this result is that the mass of a spherically symmetric black hole coupled
to scalar matter cannot be defined at the canonical level! In my opinion
this is quite worrying, especially from the quantum point of view.
Finally, a related point is that 0+1 dimensional solutions of dilaton
gravity coupled to scalar matter have no horizons -- see Refs.\
\cite{Filippov} and, for the case $V(\phi)=2/\sqrt{\phi}$, Ref.\ \cite{cd}
-- or, in other words, ``black holes have no scalar hair''.

The difficulty in the quantization of dilaton gravity coupled to a
massless scalar field is evident in any of the procedures described above.
For instance, even though the canonical transformation to free fields
described in Sect.\ \ref{dirac} can be formally implemented, the
linearization of the constraints at the basis of the Gupta-Bleuler
quantization is no longer possible. Anomalies are present and a consistent
quantization requires a modification of the theory \cite{Jackiw}.

\section*{Acknowledgements}

I am grateful to the organizers of the conference {\it Particles, Fields
\& Gravitation '98} for hospitality and financial support. I am indebted
to my friends and collaborators Vittorio de Alfaro and Alexandre T.\
Filippov for interesting discussions and useful suggestions on various
questions connected to the subject of this paper. This work has been
supported by a Human Capital and Mobility grant of the European Union,
contract no.\ ERBFMRX-CT96-0012.

\end{document}